# Evaluation of nano-frictional and mechanical properties of a novel Langmuir–Blodgett monolayer/self-assembly monolayer composite structure

Guang-hong Yang, Shu-xi Dai, Gang Cheng, Ping-yu Zhang,[a)] and Zu-liang Du[b)]
*Key Laboratory for Special Functional Materials, Henan University, Kaifeng 475001, People's Republic of China*

A novel stearic acid (SA)/3-aminopropyltrethoxysilane (APS) composite structure was fabricated using the combined method of the Langmuir–Blodgett technique and self-assembly monolayer (SAM) technique. Its frictional, adhesive properties and interface contact types between the atomic force microscope tip and the samples were evaluated based on Amonton's laws and the general Carpick's transition equation, respectively. The results showed that the tip–sample contacts corresponded to the Johnson–Kendall–Robert/Derjaguin–Muller–Toporov (DMT) transition model for $SiO_2$, APS-SAMs, and the unheated SA-APS composite structure, and for the heated SA-APS bilayer to the DMT model. Frictional forces for the four samples were linearly dependent on external loads at higher loads, and at lower loads they were significantly affected by adhesive forces. Frictional and scratching tests showed that the heated SA-APS composite structure exhibited the best lubricating properties and adhesion resistance ability, and its wear resistance capacity was greatly improved due to the binding-mode conversion from hydrogen bonds to covalent bonds. Thus, this kind of composite bilayer might be promising for applications in the lubrication of nano/microelectromechanical systems.

## I. INTRODUCTION

In fine systems such as nano/microelectromechanical systems (N/MEMSs) and computer recording, conventional tribological mechanisms become ineffective due to high surface-to-volume ratio and the greater importance of surface chemistry, adhesion, and surface structure. Microtribology and nanotribology have become an ineluctable subject.[1] But, the real tribological processes involve many physical and chemical properties at interfaces that are in relative motion; it is difficult to understand its underlying mechanism completely and clearly. In recent years, the introduction of atomic force microscopy (AFM)/friction force microscopy (FFM) presented us with the possibility of mapping both friction and topography on a solid surface at nanoscale level.[2] A "single-asperity" contact model between the tip of the atomic force microscope and the sample surface are formed, which enables us to readily understand the origin of the friction. According to macroscopic theories, friction is proportional to the external load, which matches well with experiments. But AFM studies proved that frictional and mechanical behaviors at the nanometer scale were substantially different from bulk behaviors[3] and friction as a function of load was not subject to linear Amontons' law, especially for low loads[1,3–8]; most results exhibited complicated time,[6] load,[9–11] and velocity dependencies.[7,10] Friction and wear are connected with adhesion[12] (surface energy) or adhesion energy hysteresis[8,11,13] and the breaking of adhesive junctions.

In resolving the molecular-level, friction-related problems, Langmuir–Blodgett (LB) films[6,11,14–16] and self-assembly monolayers (SAMs)[6,11,17–19] were ideal not only for modeling boundary lubrication films due to their compact and ordered structures on solid surfaces, but also for potential applications in practical nanotribological fields. LB films were first applied to lubrication studies by Langmuir. Despite the higher order and tighter molecular arrangement, the formation of LB films was mainly dependent on physical adsorption; the thermal and mechanical stability were bad, and the binding strength between film and substrate was weak. So much work[14–16] has been done to improve the thermal and mechanical stability of LB films. SAMs based on chemical bonds possess good binding forces between layers and between film and substrate; therefore, a film fabrication method that combined the LB technique with the SAMs technique was brought forward by Duschl et al.[20]

Address all correspondence to these authors.
[a)]e-mail: pingyu@henu.edu.cn
[b)]e-mail: zld@henu.edu.cn

and Wong et al.[21] To our knowledge, tribology-related research on these kinds of composite films has not yet been reported.

In this article, an ultrathin stearic acid (SA)/3-aminopropyltrethoxysilane (APS) composite structure was fabricated by a combined method using the LB technique and SAMs technique. The tip–sample contact types of the different samples were determined based on equivalent lateral contact stiffness and the derived general Carrick's equation. Their frictional and mechanical properties at the nanoscale level were evaluated using AFM/FFM systematically.

## II. THEORETICAL BACKGROUND

The tip–sample contact has vital effects on frictional phenomena at interfaces. Based on the Hertz contact theory, three contact models have been established successively. The Johnson–Kendall–Robert (JKR) model[22] describes the contact radius (area) variation under different loads when the adhesion forces outside the area of contact are neglected and elastic stresses at the contact edge are infinite. In the Derjaguin–Muller–Toporov (DMT) model,[23] adhesion forces outside the contact area are taken into account and do not cause surface deformation. Maugis[24] utilized a Dugdal potential (square well) to crudely approximate the real interaction (i.e., the Lennard–Jones potential), which can be considered to constitute a good approximation of the real contact describing the transition regime between the JKR and DMT models. But, for a given contact the appropriate model choice depends mainly on the strength and range of the adhesive force at the interfaces.[5] A transition parameter, $\lambda$, is used to determine which model is suitable for describing the actual contact. The parameter, to a certain extent, reflects the adhesion force range and the elastic deformations it causes. If $\lambda > 5$, the JKR model is valid, and if $\lambda < 0.1$, the DMT model applies. Values between 0.1 and 5 correspond to the transition case. Recently, Carpick et al.[5] also presented us another with transition parameter $\alpha$, which can be taken as a priori of the contact mechanics regime. For the contact radius $a$, the generalized equation[5] is

$$\frac{a}{a_0} = \left(\frac{\alpha + \sqrt{1 + F_n/F_{ad}}}{1 + \alpha}\right)^{2/3} \quad . \quad (1)$$

Here $a_0$ is the contact radius at zero load, $F_n$ is the normal external load, $F_{ad}$ is the adhesion force, and $\alpha$ is the transition parameter connected with the Maugis' elasticity parameter $\lambda$ according to the following equation:

$$\lambda = -0.924 \ln(1 - 1.02\alpha) \quad . \quad (2)$$

Note that $\alpha = 1$ corresponds to the JKR case and $\alpha = 0$ corresponds to the DMT case, and $1 > \alpha > 0$ corresponds to the transition regime; $a_0$ and $F_{ad}$ can be connected with the parameter $\alpha$, which has been reported on in detail in other articles.[5,23]

On the other hand, the effects of contact stiffness on adhesive properties have been discussed by different authors.[22,25,26] For a small displacement, the cantilever–tip–sample interaction can be expressed by the following equation[25]:

$$k_{tot} = \frac{dF_L}{dx} = \left(\frac{1}{k_{torsion}} + \frac{1}{k_{contact}}\right)^{-1} \quad , \quad (3)$$

where $F_L$ and $x$ are the lateral force and lateral displacement, respectively, and $K_{tot}$ and $K_{torsion}$ denote the equivalent lateral contact stiffness and the cantilever torsion constant, respectively. $K_{contact}$, the lateral contact stiffness, for the AFM tip–sample contact, is given by[25,26]

$$k_{contact} = 8G^*a \quad . \quad (4)$$

Here, $G^* = [(2 - \nu_1)/G_1 + (2 - \nu_2)/G_2]^{-1}$, where $\nu_1$, $\nu_2$, $G_1$, and $G_2$ are the respective Poisson's ratios and shear moduli of tip and sample.

## III. EXPERIMENTAL DETAILS

### A. Film preparation

Silicon wafers were cleaned by ultrasonic cleaner in deionized water and ethanol, were boiled in chloroform in turns, and were hydroxylated by immersing them in a piranha solution [i.e., a mixture of 7:3 (v/v) 98% $H_2SO_4$ and 30% $H_2O_2$] at 90 °C for 30 min. They were then fully rinsed with ultrapure water and dried in the $N_2$ flows; then they were placed into the 5 mM APS solution in a mixed solvent of acetone and ultrapure water (v/v = 5:1), and were held there for 12 h. Then, the wafers were extracted, rinsed with acetone and water, and held at 110 °C in a high-humidity atmosphere for 1 h. The APS monolayer on the Si substrate was used as the substrate for the SA LB monolayer. The composite SA-APS bilayer was fabricated with an Atemeta LB-105 trough made in Paris, France. The SA in the chloroform solution, at a concentration of $10^{-3}$ M, was spread over the deionized aqueous subphase (pH 6.5; conductance 0.1 μS/cm) and was held there for 10 min; then, the SA Langmuir monolayer was compressed at a velocity of 5 cm$^2$/min and transferred onto the APS-modified Si substrate at a surface pressure of 30 mN/m with a deposition rate of 5 mm/min. To enhance the binding strength between the film and substrate, the SA-LB bilayer was treated by heat curing at a constant temperature of 180 °C, as shown in Fig. 1.

### B. Contact angle measurement

The measurement of the static contact angle of the samples was carried out in atmospheric conditions at room temperature using a Kyowa contact-angle meter (Kyowa Interface Science Co., Ltd., Japan). At least four replicated measurements were carried out for each sample, the average static contact angles are used.

### C. AFM/FFM analysis

Our scanning probe microscope is commercially available (multifunctional SPA 400; Seiko Instruments Inc, Tokyo, Japan). All measurements were carried out using a contact mode in ambient air (relative humidity 40%; temperature 20 °C). The data on frictional forces and $K_{tot}$ in this article were collected using a triangle silicon cantilever with a $Si_3N_4$ tip; its normal and lateral spring constants were 3.5 and 25.0 N/m, respectively, according to the reported calibration method.[8] The normal sensitivity factor ($S_N$) can be obtained directly by the force curve, and the lateral sensitivity ($S_L$) is calculated[27] by frictional loops with a small scanning span. Frictional data were obtained from the subtraction between the forward and backward traces, and each of these data points was from the average of five individual data points obtained from a single loop. $K_{tot}$ was calculated according to the ratio of lateral force to lateral displacement at the original sticking part of the friction loop.[26]

The binding strength between film and substrate was evaluated using a Ti/Pt-coated silicon tip with a rectangle silicon cantilever, and its normal spring constant is 8.2 N/m based on calculations. Four squares (1 μm × 1 μm) were scratched out at increasing applied loads (410, 820, 1230, and 1640 nN) on the composite SA-APS bilayer with and without heat curing, as shown in Fig. 2. Then, a larger observation area (5 μm × 5 μm) was scanned at a velocity of 1 Hz under the lighter load (5 nN) to obtain topographical and frictional images.

## IV. RESULTS AND DISCUSSION

### A. Contact angle and morphology

The contact angle depends on several factors such as surface roughness, the manner of surface curing, cleanliness, and surface chemical composition.[28,29] Generally, the variation of the contact angle can reflect the variation of chemical composition on solid surfaces. To prove the formations of the APS SAMs and SA-APS bilayers, we obtained contact angle values for water on the sample surfaces. For the hydrophilic surface of $SiO_2$, the contact angle is <7.0° and the concordant value is about 56.6° for the APS-modified surface; this value is a little larger than that of the polyethyleneimine (PEI)-modified and APS-modified surfaces reported by Ren and colleagues[17,30]; once the SA Langmuir monolayer was transferred to the surface and the contact angle greatly increased to 101.0°, the resulting surface becomes hydrophobic due to the $CH_3$- groups. The variation of contact angle, for the three different terminal groups, accords with the relationships reported in the articles.[18,31] Figure 3 showed the topographic image [Fig. 3(a)] and the frictional image [Fig. 3(b)] of the APS monolayer. The brightness and darkness on the color scale denote the height and depth of the topography (or friction), respectively. Figure 3(a) exhibited a strictly smooth and homogeneous morphology, and the root mean square surface roughness is about 0.2 nm over the scale shown (2 μm × 2 μm), except for some drawbacks in the darker area. It can be seen from the frictional image that these drawbacks correspond to the higher frictional forces because the hydrophilic $SiO_2$ surface has higher friction than the $NH_2$-modified surface.[18,31]

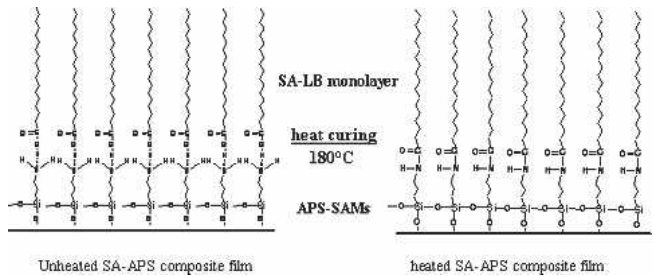

FIG. 1. Schematic diagram of the composite SA-APS bilayer; weaker hydrogen bonds were converted into covalent bonds by heat curing at 180 °C.

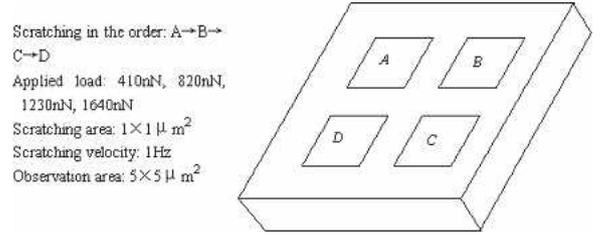

FIG. 2. A schematic graph of the nanoscratching test.

### B. Tribological behavior characterization

Equivalent lateral contact stiffness and frictional force versus normal load, for the $SiO_2$, SAMs, and SA-APS bilayers with or without heat curing, are plotted in Figs. 4, 5, 6, and 7, respectively. According to Carpick's transition equation and lateral contact-stiffness model, equivalent lateral contact stiffness has the following relationship[32]:

$$K_{tot} = \left( \frac{1}{8G^*\overline{a_0}\left(\frac{R}{K}\frac{F_n}{\overline{F_{ad}}}\right)^{1/3}} \left( \frac{1+\alpha}{\alpha+\sqrt{1+F_n/F_{ad}}} \right)^{2/3} + \frac{1}{K_{torsion}} \right)^{-1}. \quad (5)$$

Here, $\overline{a_0}$, $\overline{F_{ad}}$ are the respective parameterized values of the $a_0$ and $F_{ad}$; their correlations have been described in other articles. Lateral contact stiffness data about four samples were fitted to Eq. (5), letting the transition parameter and $8G^*\overline{a_0}\,[(R/K)(F_n/F_{ad})]^{1/3}$ be free parameters. Table I gives the fitting and calculation results. It can be seen from these results that the tip–sample contact can be described by the JKR–DMT transition case for the SiO$_2$ ($\alpha = 0.1553$), the APS monolayer ($\alpha = 0.2338$), and the unheated SA-APS bilayer ($\alpha = 0.5787$), but for the heated SA-APS bilayer, due to the transition parameter $\alpha = 0.0387$ ($\approx 0$, $\lambda < 0.1$), the tip–sample contact corresponds to the DMT model. Wei et al.[8] studied APS and octadecyltrichlorosilane (OTS) SAMs deposited on mica in air at relative humidity (HR) = 0.55, and they obtained results that were similar to ours in that the tip–sample contact was in the transition regime for the two samples. Pietrement and Troyon[32] showed the effect of adhesive forces, and found that in air the contact corresponds to the DMT model for the silicon, mica, carbon fiber, and epoxy, but that in a vacuum the contact corresponds to the JKR–DMT transition model due to the absence of capillary force. Actually, the transition parameter[5] approximately reflects the variation of the ratio of the adhesion force range to the elastic deformations at the interfaces. Thus, we must consider the effects of both adhesion force and elastic deformations at the interfaces. For the $SiO_2$, APS monolayer, and unheated SA-APS bilayer, the contact angle changes gradually, which means that the effect of the capillary force will become weaker and the spatial range over which the adhesive force acts will become shorter, so it is obvious that the increase of the transition parameter in our experiments reflected the lessening of the ratio of the force range to the deformation. But when heated the SA-APS bilayer exhibits long-range adhesive force; we think that this mainly results from the difference of the deformation between the tip–sample contacts. The elastic and plastic deformations occur much more easily for the unheated SA-APS bilayer and always recover with difficulty due to the weak molecular–molecular interaction (i.e., the hydrogen bonding). After heat curing, the hydrogen bonding was converted into the covalent bond, tight and ordered ultrathin films were formed, as shown in the following scratching test, the mechanical stability was enhanced, and these films were hardly deformed by adhesion forces; thus, we obtained the smaller transition parameter for the heated bilayer, and the tip–sample contact corresponded to the DMT model.

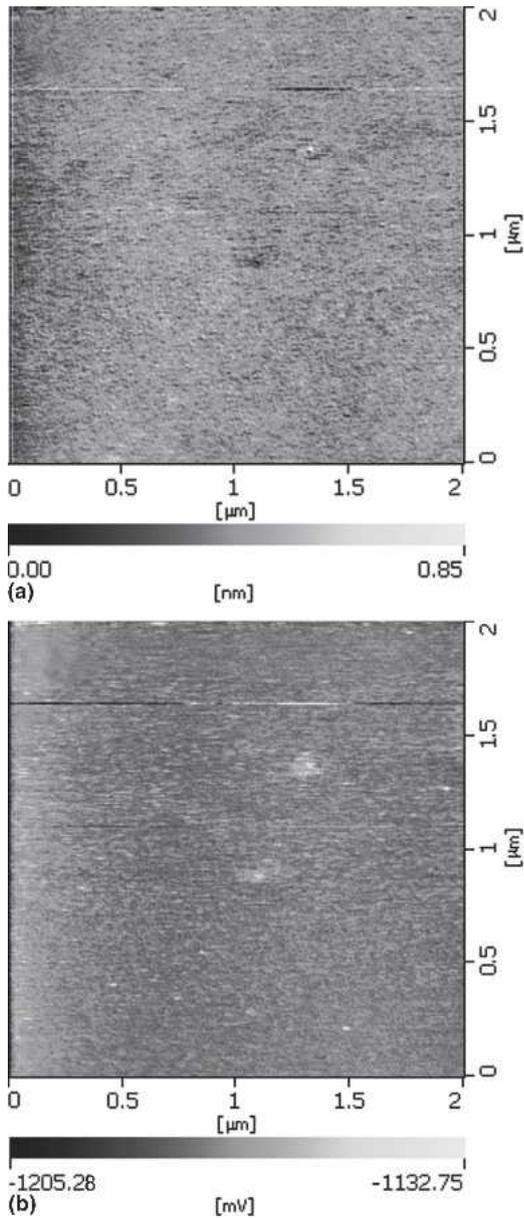

FIG. 3. (a) Topographic and (b) frictional images of the APS monolayer. The brightness and darkness on the color scale denote the height and depth of the topography (or friction), respectively.

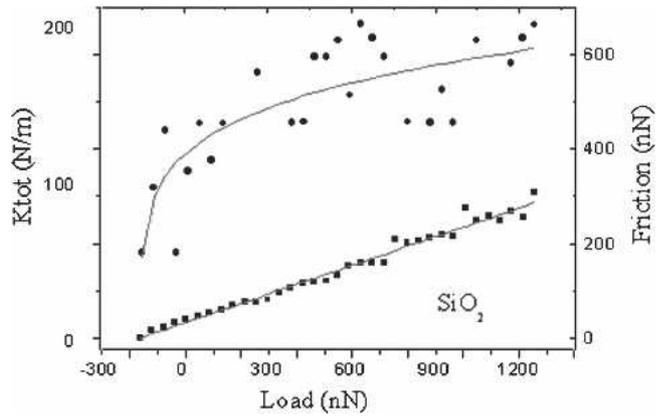

FIG. 4. Equivalent $K_{tot}$ and frictional force versus normal loads for SiO$_2$. The data points using a rectangular lever in the scratching tests for $K_{tot}$ are fitted with Eq. (5), and frictional force is linearly proportional to the load.

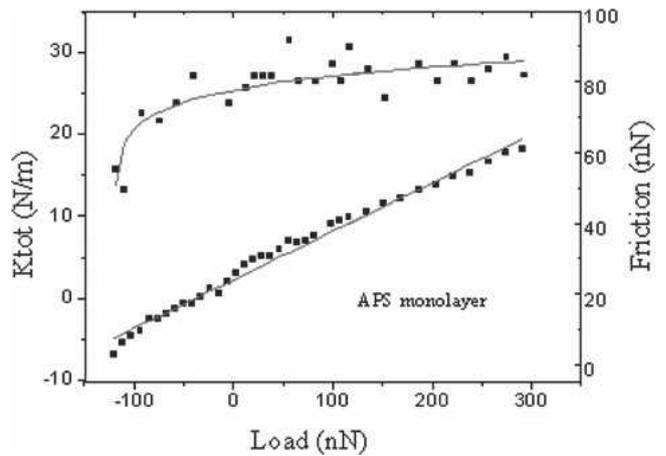

FIG. 5. Equivalent $K_{tot}$ and frictional force versus normal loads for the APS monolayer. The data points for $K_{tot}$ are fitted with Eq. (5), and the frictional forces are linearly proportional to the load and are affected by the adhesion force at low loads.

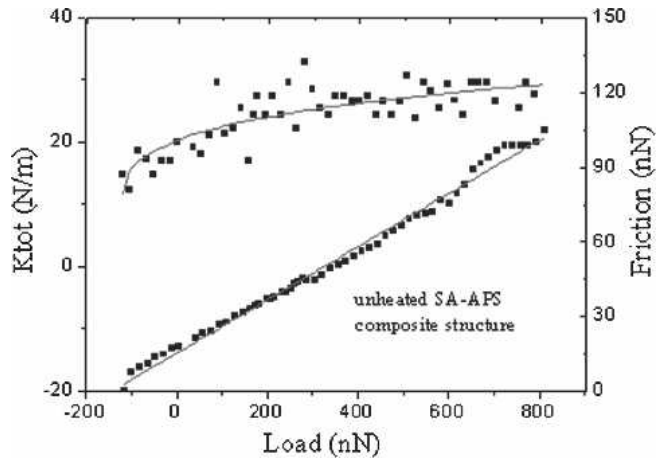

FIG. 6. Equivalent $K_{tot}$ and frictional force versus normal loads for the unheated SA-APS composite structure. The data points for $K_{tot}$ are fitted with Eq. (5), and the frictional force is linearly proportional to the load.

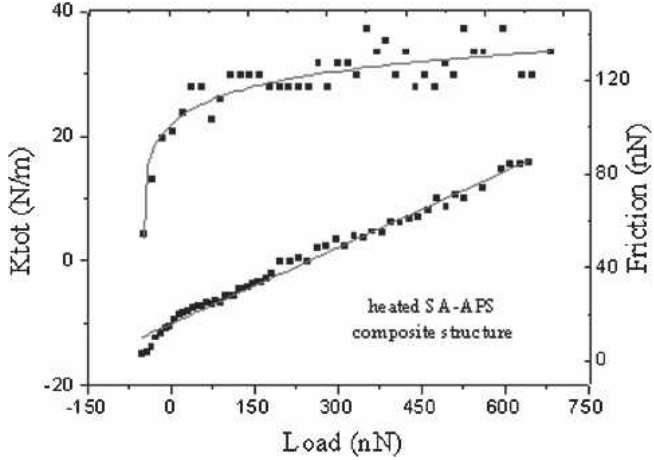

FIG. 7. Equivalent $K_{tot}$ and frictional force versus normal loads for the heated SA-APS composite structure. The data points for $K_{tot}$ are fitted with Eq. (5), and the frictional force is linearly proportional to the load, except for a deviation from linearity at low loads.

TABLE I. Interface contact properties deduced from lateral contact stiffness fitted with Eq. (5) for the four studied samples.

| Data | $SiO_2$ | SAM | Unheated SA-APS bilayer | Heated SA-APS bilayer |
| --- | --- | --- | --- | --- |
| $\alpha$ | 0.1553 | 0.2338 | 0.5787 | 0.0378 |
| $\lambda$ | 0.1568 | 0.2517 | 0.8248 | 0.03633 |
| $F_{ad}$ (nN) | 184 | 120 | 119 | 48.7 |
| $\gamma$ (J/m2) | 1.2088 | 0.7561 | 0.7786 | 0.3495 |
| $\mu$ (cof) | 0.2034 | 0.1372 | 0.1069 | 0.1029 |

From these interface energies ($\gamma$, also considered as works of adhesion) and the data in Table I, we found that the interface energies had the same variation trends as adhesive forces. Interface energy and adhesive force were the highest for the tip–$SiO_2$ interface. For the tip–APS interface and the tip-unheated SA-APS bilayer, adhesion energies and adhesive forces occurred in the same order. Ren and colleagues[17,30] thought that capillary interactions play important roles in interface adhesion, and adhesion forces at the interfaces included capillary forces as well as solid–solid interactions, which consisted of van der Waals forces, electrostatic forces, and chemical forces. Capillary interactions, which are closely related to surface wettability, were the main contributions to adhesion energies. Electrostatic force was usually negligible, and van der Waals forces made a small contribution. Chemical forces might play an important role at interfaces, such as the $Si_3N_4$ tip and $SiO_2$ interface. It was easily understood that the highest interface energies for tip–$SiO_2$ contact were mainly attributed to combined contributions of capillary interactions and interface-mated chemical forces. The second highest energies at the tip–APS interface might result from the weakened capillary interactions and van der Waals forces due to molecular–molecular adsorptions at the interfaces because tip scanning might adsorb some organic molecular. But for the tip and the unheated SA-APS cooperation, the interface energy was almost equal to that of tip–APS contact, we thought that van der Waals forces were the main contributors because the binding strength is weaker between SA and the amino-modified surface, and the molecular-chain length and its movability significantly affected deformation; thus, the real contact area was relatively large and at the interface there were higher van der Waals forces. For the heated SA-APS bilayer, the adhesion energy was the lowest; this could also be interpreted by molecular-chain stability and the small real contact area. In the later discussion, we knew that there were many microasperities, which were formed by the release of adsorbed water during the heat process on the surface of SA-APS bilayer. These asperities made the surface roughness increase, and, furthermore, the hydrophobicity of the surface was enhanced and adhesion energies were reduced.[29]

The variations of frictional forces versus loads were also shown in Figs. 4–7. For the four samples, a linear dependence between frictional force and the external

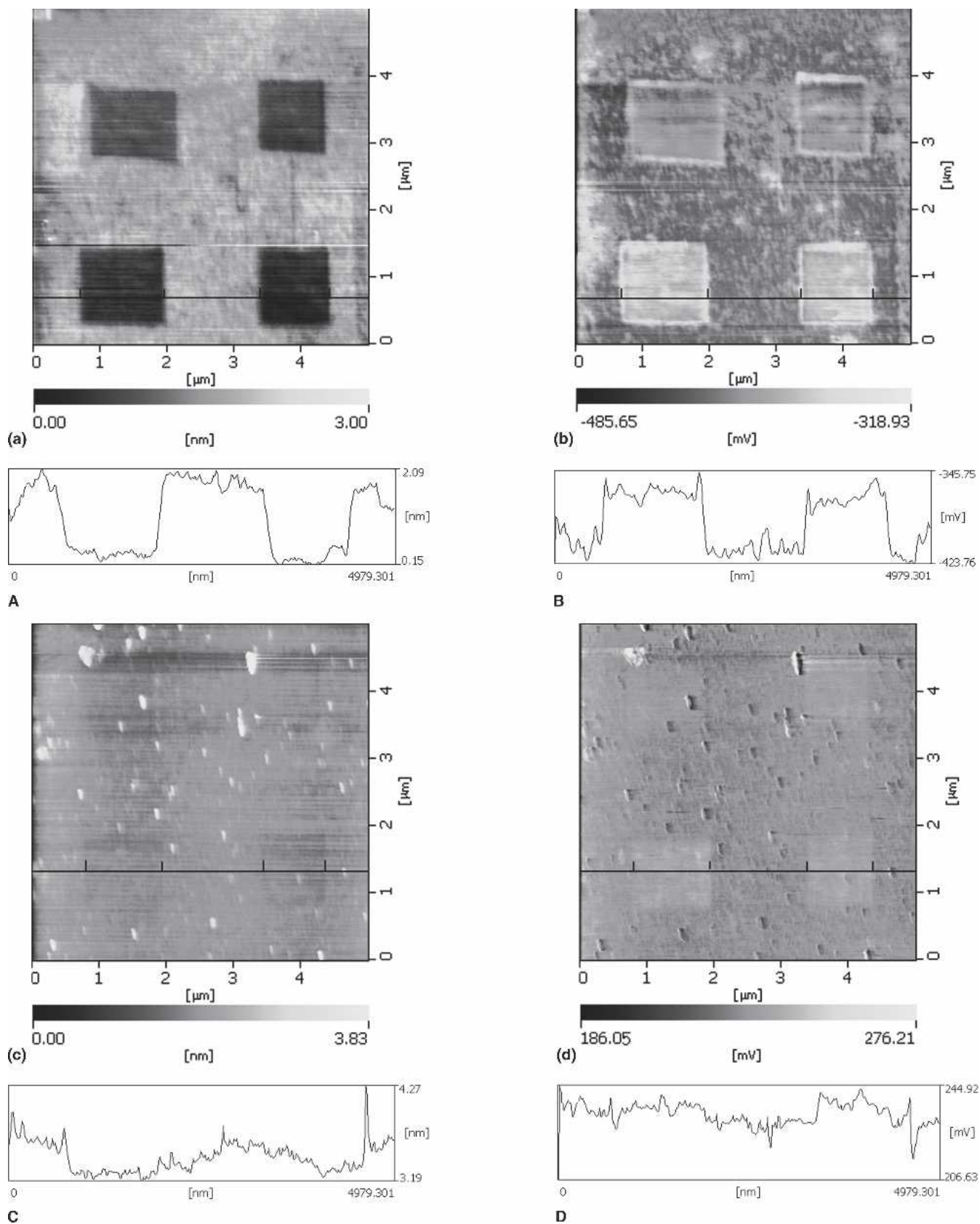

FIG. 8. (a, b) Results of nanoscratching tests for the SA-APS bilayer without heat curing and (c, d) with heat curing. (a, c) are respective topographical images of the scratching area; (b, d) and are corresponding frictional images. Curves A–D are the respective profiles corresponding to Figs. 8(a)–8(d).

2434

load could be clearly found, except for small deviation at the low loads, which was mainly attributed to the effects of adhesive forces. We obtained our frictional coefficient data based on the linear Amonton's law and found a friction coefficient of 0.2034 for the $SiO_2$ surface, which is an intermediate value between those given by Ruan and colleagues[17,30] (0.07) and by Meyer (0.3).[32] Recently, Pietrement and Troyon[32] obtained values of 0.15 in air and 0.14 in a vacuum, but our measurements were close to the macroscale friction coefficient given by Bhushan[32] (0.18). Friction coefficients for the APS monolayer and the unheated or heated SA-APS bilayer were 0.1372 and 0.1069, respectively. It was obvious that different values for the friction coefficient were concerned with many factors such as the calibration of cantilever force constants, the geometry of atomic force microscope tips and surface damage. Many groups have reported relationships between friction and adhesion force or adhesion hysteresis. Our results showed that friction has the same variation trends as adhesion forces.

It was found that both the unheated and heated SA-APS bilayers showed excellent lubricating properties, which could be mainly attributed to the ordered SA-LB monolayer. That is to say, the ordered long chains of the SA monolayer with one end attached to the substrate surface had a great freedom to swing and rearrange along the sliding direction under shear forces, and yielded a smaller resistance. As for the relatively large amount of friction on the amino-modified surface, it could be due to higher adhesion and more severe molecular disordering. The friction force for the unheated SA-APS film rapidly increased as external loads exceeds 600 nN and became a plane as the load reached 700 nN, which was thought to be the results of the wear of the SA-LB monolayer. Friction coefficients were almost equivalent for unheated or heated films, except that the slightly larger value for the unheated film due to the sliding process could involve the damage of hydrogen bonds and the transfer of molecular chains.

### C. Nanomechanical properties

It was important for the application of nanolubricating films to devices to own not only these lubricating properties, but also wear-resistance capacity. Figures 8(a) to 8(d) illustrated our scratch test results of the unheated and heated SA-APS composite structures under different external loads; Figs. 8(a)–8(d) were the corresponding topography and frictional profiles. Before being heated, the binding strength between the SA monolayer and APS-modified substrate was weaker, and the SA monolayer was easily compressed more tightly or scratched off due to the damage to the hydrogen bonds; the depth of the scratches increased with enhanced of the external loads. Brostow et al.[33] reported detailed scratching results by the addition of a fluorinated poly (aryl ether ketone) to a commercial epoxy resin and curing at either 24 or 70 °C. Shojiro et al.[31] and Wang et al.[34] studied the tribological characteristics of perfluoropolyether with and without heat curing, respectively. They thought that these freely replenished lubricants had roles in reducing friction and improving antiwear properties. But, our results indicated that the scratched areas showed relatively greater friction than those unscratched areas, as shown in Figs. 8(a)–8(d) (topography and frictional profiles). We thought that, on the one hand, molecular-chain transfers (or removal) can consume some energy to a certain extent; on the other hand, the scratched areas had greater contact stiffness due to the removal of SA molecules. After being heated, many small asperities, which were formed by the release of water at high temperature, appeared on the surfaces and existed at the SA and APS interfaces. On the other hand, wear depth did not change distinctly at loads ranging from 410 to 1640 nN, and the heated composite SA-APS film showed excellent wear-resistance properties, which was valuable for application to nanodevices. We also noticed slightly higher friction over the scratched area in Fig. 8(d), which was thought to be induced by the tighter molecular arrangement.[30]

### V. CONCLUSION AND REMARKS

In this article, a novel SA-APS composite structure was fabricated using a combined method using the LB and SAM techniques. Their nanotribological properties were investigated based on the derived Carpicks' transition equation. The results showed that the tip–sample contacts corresponded to the JKR–DMT transition model for $SiO_2$, APS-SAMs, and the unheated SA-APS composite structure, and for the heated SA-APS bilayer to the DMT model. Frictional and scratching tests showed that the heated SA-APS composite structure exhibited great improvement in antiadhesion and wear-resistance capacity. This indicates that this kind of composite bilayer might be promising for applications in the lubrication of N/MEMSs.

### ACKNOWLEDGMENTS

This work was supported by National Natural Science Foundation of China (Grant Nos. 90306010, 20371015, 20571024) and the Program for New Century Excellent Talents in University.